\documentclass[12pt]{article}

\usepackage[utf8]{inputenc}
\usepackage{amsmath}
\usepackage{graphicx}
\usepackage{svg}
\usepackage{float}
\usepackage{subcaption} 
\usepackage{textcomp}
\usepackage{caption}
\usepackage{siunitx}
\usepackage{authblk}
\usepackage[T1]{fontenc}
\usepackage{csquotes}
\usepackage{hyperref}
\usepackage{xurl}      
\usepackage{microtype} 
\usepackage{textgreek}

\usepackage[
  backend=biber,
  style=numeric-comp,
  giveninits=true,
  maxnames=99,
  minnames=1,
  doi=true,
  url=true,
  eprint=true,
  sorting=none,
  sortcites=false,
  bibencoding=utf8
]{biblatex}

\DeclareNameAlias{default}{family-given}

\DeclareFieldFormat{labelnumber}{\mkbibemph{#1}}

\DeclareFieldFormat{labelnumberwidth}{\mkbibparens{\mkbibemph{#1}}}

\DeclareDelimFormat[bib,biblist]{nametitledelim}{\addcolon\space}
\DeclareDelimFormat{multicitedelim}{\addcomma\space}
\DeclareDelimFormat{compcitedelim}{\textendash}

\DeclareCiteCommand{\cite}[\mkbibparens]
  {\usebibmacro{prenote}}
  {\mkbibemph{\usebibmacro{citeindex}%
   \usebibmacro{cite}}}
  {\multicitedelim}
  {\usebibmacro{postnote}}

\renewbibmacro*{cite}{%
  \printtext[bibhyperref]{%
    \mkbibemph{\printfield{labelprefix}%
    \printfield{labelnumber}%
    \ifbool{bbx:subentry}
      {\printfield{entrysetcount}}
      {}}}}

\DeclareCiteCommand{\citet}
  {\usebibmacro{prenote}}
  {\mkbibemph{\usebibmacro{citeindex}%
   \usebibmacro{cite}}}
  {\multicitedelim}
  {\usebibmacro{postnote}}

\DeclareFieldFormat{pages}{#1}
\DeclareFieldFormat{page}{#1}

\newbibmacro*{pagesoreid}{%
  \iffieldundef{eid}{%
    \iffieldundef{pages}{}{%
      \printfield{pages}%
    }%
  }{%
    \printfield{eid}%
  }%
}

\DeclareFieldFormat{journaltitle}{\mkbibemph{#1}}
\DeclareFieldFormat[article,incollection,techreport]{title}{\mkbibemph{#1}}

\DeclareFieldFormat{doi}{%
  \iffieldundef{doi}{}{%
    \addperiod\space doi:\space
    \href{https://doi.org/\strfield{doi}}{\nolinkurl{\strfield{doi}}}%
  }%
}
\DeclareFieldFormat{eprint}{%
  \iffieldundef{eprint}{}{%
    \addperiod\space
    \iffieldundef{eprinttype}{%
      \href{https://arxiv.org/abs/\strfield{eprint}}{\nolinkurl{\strfield{eprint}}}%
    }{%
      \printtext{\strfield{eprinttype}\addcolon\space\href{https://arxiv.org/abs/\strfield{eprint}}{\nolinkurl{\strfield{eprint}}}}%
    }%
  }%
}

\DeclareBibliographyDriver{article}{%
  \usebibmacro{bibindex}%
  \usebibmacro{begentry}%
  \printnames{author}\addperiod\space
  \printfield[title]{title}\addperiod\space
  \printfield{journaltitle}\addspace
  \iffieldundef{year}{}{\mkbibbold{\printfield{year}}\addcomma\space}%
  \iffieldundef{volume}{}{%
    \mkbibemph{\printfield{volume}}\addcomma\space
  }%
  \usebibmacro{pagesoreid}%
  \printfield{doi}%
  \printfield{eprint}%
  \usebibmacro{finentry}%
}

\DeclareBibliographyDriver{book}{%
  \usebibmacro{bibindex}%
  \usebibmacro{begentry}%
  \printnames{author}\addperiod\space
  \printfield[title]{title}\addperiod\space
  \printlist{publisher}\addcomma\space
  \printlist{location}\addcomma\space
  \printfield{year}\addperiod\space
  \usebibmacro{finentry}%
}

\DeclareBibliographyDriver{incollection}{%
  \usebibmacro{bibindex}%
  \usebibmacro{begentry}%
  \printnames{author}\addperiod\space
  \printfield[title]{title}\addperiod\space
  \printtext{In: }\printfield{booktitle}%
  \iffieldundef{pages}{}{, \printfield{pages}}%
  \addcomma\space\printfield{year}\addperiod\space
  \printfield{doi}%
  \usebibmacro{finentry}%
}

\usepackage{etoolbox}
\makeatletter
\patchcmd{\@maketitle}
  {\ifx\@empty\@dedicatory}
  {\vspace{-3em}\ifx\@empty\@dedicatory}
  {}{}
\makeatother

\title{\textbf{Molecular Dynamics Simulations of Collision Cascades in Niobium:}\\ Comparing Interatomic Potentials}
\author[a]{S.~Mondal\thanks{Corresponding author. Email: \href{mailto:smondal@barc.gov.in}{smondal@barc.gov.in}}}

\author[a]{U.~Bhardwaj}
\author[a]{A.~Majalee}
\author[a,b]{V.~Mishra}
\author[a,b]{M.~Warrier}

\affil[a]{Computational Analysis Division, BARC, Vizag, AP, India - 531 012}
\affil[b]{Homi Bhabha National Institute, Anushaktinagar, Mumbai, Maharashtra, India - 400 094}

\date{}

\begin{document}
\maketitle

\begin{abstract}
Radiation damage in structural materials is a major challenge for advanced nuclear energy systems, and niobium is of particular interest due to its high melting point, mechanical strength, and corrosion resistance. To better understand its radiation response, we carried out large-scale molecular dynamics simulations of collision cascades in pure niobium at 300~K over a primary knock-on atom (PKA) energy range of 1--75~keV, employing four interatomic potentials: an embedded atom method (EAM), two Finnis--Sinclair models (FS-1 and FS-2), and a machine learning--based spectral neighbor analysis potential (SNAP) we developed. All reproduce the general features of cascade formation but differ significantly in defect production, clustering, and morphology. At low energies, defect generation follows trends governed by threshold displacement energy (TDE) and the stiffness-to-range ratio ($|S/R|$). At higher energies, subcascade formation makes defect evolution dependent on the combined effects of $|S/R|$, average TDE, and other material-specific factors. Vacancy clustering dominates over interstitial clustering across all cases: EAM produces the largest vacancy clusters and the highest clustering fraction, while SNAP shows the strongest interstitial clustering. Morphological analysis indicates that EAM forms a balanced mix of $1/2\langle 111 \rangle$, $1/2\langle 110 \rangle$ loops, C15 rings, and hybrid structures; FS-2 favors extended $1/2\langle 111 \rangle$ dumbbells, crowdions, and dislocation loops; whereas FS-1 and SNAP generate more compact or disordered clusters, with SNAP produces a high fraction of C15-like rings (maximum size up to nine atoms) that may evolve into dislocation loops of $1/2\langle 111 \rangle$ and $\langle 100 \rangle$. These findings give clear insights into how niobium reacts when exposed to irradiation, especially at high energies.
\end{abstract}


\section{Introduction}
The performance and lifespan of nuclear reactors depend heavily on how well materials withstand extreme radiation environments~\cite{busby2008materialsdegradation,b,a3}. When energetic particles strike these materials, they trigger a series of atomic collisions~\cite{Stoller2012_PrimaryDamage,NORDLUND2018450}. This process creates point defects, clusters of defects, and other structural changes at the atomic level~\cite{Stoller2012_PrimaryDamage,NORDLUND2018450}. Gaining a clear understanding of this initial damage, caused by collision cascades and the resulting structural changes, is essential for predicting how the material's microstructure will evolve over time~\cite{YOSHIIE1999296}. As radiation damage accumulates, defects migrate, combine, and may trigger phase transformations, which gradually alter the bulk properties of materials. These radiation-induced changes—such as hardening, embrittlement, swelling, and dimensional instability—strongly influence the mechanical performance and structural integrity of reactor components~\cite{zinkle2012_microstructure,aitkaliyeva2017irradiation}.

Niobium is a refractory metal with a body-centered cubic (BCC) structure and is well known for its excellent performance at high temperatures~\cite{r1,goldberg1972niobium}. Niobium-based alloys are valued for their good thermal stability, high thermal conductivity, and compatibility with liquid metal coolants~\cite{r2,zinkle2000operating}. One of niobium’s key advantages is its low thermal neutron capture cross-section~\cite{Nakamura02112023}, which makes it a useful alloying element in nuclear materials, especially in zirconium-based alloys. Even small amounts of niobium can enhance the mechanical strength of zirconium alloys~\cite{met14060646}. Because of these properties, niobium is considered a promising material for use in advanced nuclear reactors and future fusion energy systems~\cite{goldberg1972niobium,r2,zinkle2000operating}.\\

Radiation-induced microstructural changes in pure niobium have been the subject of extensive experimental research over the years. One notable study by Tucker \textit{et al.}~\cite{Tucker01091967} used transmission electron microscopy (TEM) to examine neutron-irradiated niobium. They reported the presence of dislocation loops with Burgers vectors of $\frac{1}{2}\langle111\rangle$, $\frac{1}{2}\langle110\rangle$, and $\frac{1}{2}\langle100\rangle$, and estimated a dislocation density of approximately $5 \times 10^{15} \, \mathrm{cm^{-3}}$ at a dose of $2 \times 10^{18} \, \mathrm{n/cm^2}$. Further insights were provided by Brimhall \textit{et al.}~\cite{Brimhall01011973}, who investigated high-purity niobium irradiated with 7.5 MeV tantalum ions at elevated temperatures of 800\textdegree{}C and 900\textdegree{}C. Their TEM analysis revealed that void formation intensified with increasing dose, showing a strong correlation with radiation exposure up to about 100 displacements per atom (dpa). Beyond this dose, void swelling saturated and a body-centered cubic (BCC) void lattice structure emerged. They also highlighted the significant role of temperature in determining both the degree of swelling and the organization of voids, with higher temperatures promoting more structured void lattices and greater swelling. More recently, Dutta \textit{et al.}~\cite{dutta2018microstructural} investigated the microstructural evolution of niobium under 5 MeV proton irradiation at doses ranging from $5.4 \times 10^{16}$ to $1.2 \times 10^{18}$~p/cm$^2$ (0.01–0.14 dpa). Using X-ray diffraction line profile analysis (XRDLPA) and transmission electron microscopy (TEM), they observed a sharp initial reduction in domain size due to the formation of defect clusters. At higher doses, this reduction leveled off, indicating a saturation effect. Additionally, they reported a rise in microstrain and dislocation density at lower doses, which decreased at higher irradiation levels. TEM images suggested that this behavior resulted from the evolution of defect clusters into dislocation loops, though the specific types of loops were not identified in their study.\\

While experimental studies have provided valuable insights, computational methods such as Molecular Dynamics (MD) simulations offer a powerful tool for analyzing atomic-level radiation effects~\cite{Stoller2012_PrimaryDamage,NORDLUND2018450,PHYTHIAN1995245,Warrier01082013}. However, MD investigations specifically focused on pure niobium remain limited. De-Ye Lin \textit{et al.}~\cite{lin2017molecular} performed MD simulations at PKA energies of 1, 5, 10, and 20~keV across temperatures of 300~K, 600~K, and 900~K. They studied cascade evolution, defect generation, clustering behavior, and temperature effects. Their findings showed a low fraction of defect clusters, which further decreased with increasing temperature. Additionally, the cluster sizes observed were small. They noted that defect production in niobium was higher than in tungsten, but similar to vanadium and molybdenum in terms of trends and cluster characteristics. Although this study offered useful insights, it also has certain limitations. The PKA energies considered (1--20~keV) are relatively low, leaving open the question of how niobium responds to higher-energy cascades. Moreover, the work did not explore the detailed morphology of defects, such as the presence of $\tfrac{1}{2}\langle 111\rangle$ dumbbells, $\langle 100\rangle$ dumbbells, crowdions, dislocation loops, C15-type rings and its basis~\cite{BHARDWAJ2021110474}, or other extended defect structures.\\

In this study, we employ MD simulations to explore radiation effects in pure niobium at higher PKA energies, ranging from 1~keV to 75~keV, at a temperature of 300~K. We present a detailed analysis of Frenkel pair evolution, defect count distributions, subcascade formation energies, defect clustering fractions, and the morphological characteristics of defect clusters. Comparisons are made using a Machine Learning SNAP, EAM and FS interatomic potentials to understand how potential choice influences defect formation and clustering behavior.

\section{Interatomic Potentials Description}

This study uses four interatomic potentials to investigate primary radiation damage in niobium: a machine learning-based Spectral Neighbor Analysis Potential (SNAP), an Embedded Atom Method (EAM) potential, and two Finnis–Sinclair variants, FS-1 and FS-2.\\

 We developed the SNAP potential~\cite{Bhardwaj_2025} and trained it using a wide range of density functional theory (DFT) data, including strained structures, defects, and liquid states. To handle energetic collisions, we added a short-range correction based on the Ziegler–Biersack–Littmark (ZBL) potential~\cite{ziegler1985stopping}. The SNAP model accurately reproduces key properties of niobium, such as lattice constants, elastic behavior, and defect formation energies. It also predicts the \textless111\textgreater\ dumbbell as the most stable self-interstitial atom (SIA), which agrees with DFT results. In comparison, traditional EAM and FS potentials usually prefer the less accurate \textless110\textgreater\ configuration. The EAM potential  was developed by Fellinger et al.~\cite{PhysRevB.81.144119}. It was parameterized to reproduce the key mechanical and thermodynamic properties of niobium. To better capture the strong repulsion at short interatomic distances, we stiffened the pair interaction term with the ZBL potential in the range of 0.75–1.25 Å. This ensured a smooth transition into the repulsive regime without affecting the bulk properties. After this modification, we validated the interatomic potential (IAP) against equilibrium properties such as elastic constants, melting point, and vacancy formation energy to confirm its accuracy.\\
The FS-1 potential, originally developed by Mendelev et al.~\cite{zhang2016experimental} for Ni–Nb alloys, includes parameters suitable for pure niobium. This model already incorporates a short-range repulsive term and does not require additional ZBL blending. The FS-2 potential, proposed by Qiu et al.~\cite{Qiu_2023}, was developed for multi-component alloys but contains updated parameters specifically tuned for niobium. It also includes a ZBL correction for short-range interactions and shows good agreement with DFT predictions for defect formation and migration energies, as well as the correct dumbbell configuration for SIAs.\\

We compared the ability of the different interatomic potentials to reproduce key material properties. 
The results are summarized in Table~\ref{tab:Nb_properties}, which lists physical properties and defect energetics from each potential together with available DFT and experimental data. To evaluate the repulsive behavior of the interatomic potentials under energetic collisions, we also examined their short-range energy curves. Figure~\ref{fig:short_range} shows these curves, with stars marking the distance $R$ where the potential energy reaches 30~eV. The slope at this point gives the stiffness ($S$) of the potential, and the ratio $S/R$ indicates the relative strength of short-range repulsion. A larger $S/R$ means a stiffer potential with stronger resistance to atomic overlap, while a smaller value reflects a softer repulsion, affecting how collision cascades and radiation damage evolve~\cite{BYGGMASTAR2018530,SAND2016119}. Among the four potentials, EAM shows the largest \(|S/R|\), reflecting the strongest short-range repulsion. SNAP has the smallest \(|S/R|\), indicating the softest repulsion. FS-1 and FS-2 fall in between, with FS-2 slightly higher than FS-1. These differences in the \(S/R\) ratio may help explain variations in defect formation, cascade morphology, and defect clustering during irradiation~\cite{BYGGMASTAR2018530,}.

\section{Methodology}
We used molecular dynamics, as implemented in the LAMMPS software~\cite{THOMPSON2022108171}, to model how pure niobium responds to primary radiation damage. Cubic simulation boxes were created with dimensions provided in Table~\ref{tab:box}, and each system was equilibrated for 10\,ps using an NPT ensemble at 300\,K and zero pressure. Periodic boundary conditions were applied in all directions, and a time step of 1\,fs was used throughout. The box size was chosen to ensure that the entire collision cascade remained confined to the central region, excluding the outermost three unit cells in each direction. This was verified post-simulation, and only minimal cascade channeling effects were observed. The primary knock-on atom (PKA) was selected from atoms located near the center of the simulation box. The PKA was launched in 25 different random directions. Each simulation was run for 20~ps at 300~K. Electronic stopping effects were included following the approach described in~\cite{hemani2020inclusionvalidationelectronicstopping}. Outermost unit cells are fixed so that there is no net momentum to the system due to the energetic PKA. A width of two unit cells just within the outermost unit cells were coupled to a Berendsen thermostat at 300~K~\cite{berendsen1984molecular}. Variable time stepping was enabled to handle fast atomic movements during the early high-energy phase of the cascade. Atomic coordinates, velocities, and large displacements were tracked during the simulation. Full XYZ coordinate data and per-atom energy values were saved for detailed post-processing.\\

To analyze the large number of simulations, we used the CSaransh post-processor~\cite{BHARDWAJ2021110474,article,BHARDWAJ2020109364,Bhardwaj_2024}. CSaransh can efficiently handle large MD datasets and identify radiation-induced point defects, their clusters, and the overall damage morphology. It uses atomic position data (XYZ format) from the final simulation step to detect and classify defects based on their type, spatial arrangement, and local geometry. This allowed us to systematically assess defect structures and their evolution across different PKA energies and interatomic potentials.\\

\section{Results}
\subsection{Cascade Evolution}

Figure~\ref{fig:potential_compare} presents the time evolution of the average number of defects in pure Nb for two PKA energies, 10 and 50~keV, using four different interatomic potentials (SNAP, EAM, FS-1, FS-2). The peak in defect production occurs within about 0.8~ps for 10~keV and 3.6~ps for 50~keV. As expected, higher PKA energy leads to a later peak time, a larger peak number of defects, and a higher number of surviving defects, consistent with earlier studies on BCC metals~\cite{Stoller2012_PrimaryDamage,sun2024molecular}. In all cases, the potentials show a similar overall behavior: a sharp increase in FP production during the ballistic phase, followed by recombination and eventual stabilization. However, the predicted numbers of surviving defects differ substantially among the potentials. At 10~keV, the EAM potential predicts the highest average number of defects during both the thermal spike and the stabilized stage. FS-2 reaches a peak level close to EAM but leaves significantly fewer surviving defects, indicating more efficient recombination or different cascade dynamics. FS-1 produces slightly fewer peak defects than EAM, with a surviving defect number between those of EAM and FS-2. SNAP consistently predicts the lowest peak defect count and the lowest surviving defect number, though the latter is very close to FS-2. At 50~keV, the trends shift. FS-1 predicts the highest peak defect number, whereas EAM, FS-2, and SNAP produce nearly the same peak value. In terms of surviving defects, EAM again yields the largest number, followed closely by FS-1. SNAP and FS-2 predict similar and comparatively lower surviving defect numbers.\\

Figure~\ref{fig:defect_evolution} shows the spatial and temporal evolution of point defects during a representative 50~keV PKA cascade for SNAP, though similar behavior is also observed for EAM, FS-1, and FS-2. At 0.6~ps, a dense cloud of vacancies and interstitials appears, corresponding to the formation of a large number of Frenkel pairs. By 13.2~ps, many of these defects have recombined, leaving behind a reduced defect population. At 20~ps, the system reaches a stabilized state with vacancies and interstitials that survive because they are too far apart to annihilate directly and can only evolve through slower diffusion processes~\cite{r54}. Together, Figures~\ref{fig:potential_compare} and \ref{fig:defect_evolution} demonstrate the general mechanism of cascade evolution in niobium: rapid defect production, followed by partial recombination, and stabilization into a residual defect population.

\subsection{Subcascade Formation with Increasing PKA Energy}

To analyze cascade morphology, we determined the number of subcascades using the DBSCAN density-based clustering method~\cite{Bhardwaj_2024}, which groups nearby atomic displacements into clusters based on their spatial density. As shown in Figure~\ref{fig:subcascades}, all four interatomic potentials produce only a single cascade at low energies (1–10~keV), suggesting that the damage remains compact in this range. Around 20~keV, the number of subcascades begins to increase, indicating that the cascade starts to split into separate regions. This trend continues with increasing energy, showing more fragmentation at higher PKA values. At \SI{75}{\kilo\electronvolt}, the SNAP potential produces the highest number of subcascades, FS-1 and FS-2 give nearly similar results, while EAM shows the lowest. These differences indicate that the way each potential transfers energy affects both the spatial spread and the overall morphology of the cascade.

\subsection{Energy Dependence of Defect Production}

The number of surviving Frenkel pairs increases steadily with PKA energy, as shown in Figure~\ref{fig:powerlaw_plot}, and follows the power-law relation \(N_{\text{FP}} = a \cdot E^b\). Similar scaling has been reported in other bcc metals such as Fe, Cr, Mo, and W, where distinct exponents describe the low- and high-energy regimes~\cite{Setyawan_2015}. For vanadium, for example, exponents of 0.76 and 0.97 were observed across two energy ranges, with a morphological transition occurring near 10~keV~\cite{QIU2023101394}. In niobium, our results also reveal a clear two-regime behavior. The fitted parameters are summarized in Table~\ref{tab:powerlaw_fit}. In the low-energy range (1–20~keV), the exponent \(b_1\) varies from 0.683 (EAM) to 0.864 (SNAP). At higher energies (20–75~keV), the exponent \(b_2\) increases, ranging from 0.841 to 0.930. This increase indicates a more efficient defect production process as the PKA energy rises. The transition near 20~keV likely reflects a change in cascade morphology—from compact cascades at low energies to more extended or subcascade-dominated structures at higher energies. This interpretation is consistent with the subcascade analysis shown in Figure~\ref{fig:subcascades}, where the number of subcascades begins to rise around 20~keV for all potentials. The agreement between the power-law exponents and subcascade trends strongly supports the conclusion that subcascade formation is the underlying mechanism driving the change in defect production behavior at higher energies.

\subsection{In-cascade Clustering Behavior}

Figures~\ref{fig:interstitial_clusters} and~\ref{fig:vacancy_clusters} show the percentage of interstitials and vacancies that form clusters after collision cascades at different PKA energies from the four different interatomic potentials. These results provide insight into how defects organize themselves immediately after cascade formation. For interstitials (Figure~\ref{fig:interstitial_clusters}), clustering increases with PKA energy across all potentials. At low energies (around 1\,keV), clustering is relatively low—usually under 25\%—because fewer defects are produced and cascades are more dispersed. As energy increases, the clustering fraction rises, especially for the SNAP and FS-2 potentials, which reach values above 50\% by 20\,keV. Beyond that point, the increase slows or plateaus. SNAP consistently produces the highest interstitial clustering. This seems to result from the formation of compact, low-density cascades, as reflected by its lower $\lvert S/R \rvert$ ratio compared to the other potentials. FS-2 also performs similarly well. In contrast, FS-1 and EAM predict lower interstitial clustering across the energy range, with EAM showing more variation. In the case of vacancies (Figure~\ref{fig:vacancy_clusters}), clustering levels are much higher overall. Even at low energies, a significant fraction of vacancies form clusters—typically over 70\% for most potentials. EAM shows the highest vacancy clustering, remaining above 85\% across all energies. FS-1 also predicts strong clustering (around 75–80\%), while SNAP and FS-2 show lower values at the lowest energies but converge to 70–75\% by 20~keV. Unlike interstitials, vacancy clustering quickly reaches a high level and then remains fairly constant, indicating that vacancy aggregation happens early and is less affected by further increases in energy. Overall, these trends show that vacancy clustering is robust across all energies and potentials, while interstitial clustering is more sensitive to both the cascade energy and the choice of interatomic potential.

\subsection{Maximum Cluster Size as a Function of PKA Energy}

Figure~\ref{fig:max_inter_cluster} shows that the maximum size of interstitial clusters increases with energy for all potentials. Among them, FS-2 consistently results in the largest interstitial clusters, especially at higher energies. At 75\,keV, FS-2 forms clusters exceeding 9 interstitials, while the other potentials remain below this value. In Figure~\ref{fig:max_vac_cluster}, a similar trend is observed for vacancy clusters, with sizes generally growing with energy. The EAM potential leads to the largest vacancy clusters across most energies, reaching over 60 vacancies at 75\,keV. FS-1 shows intermediate behavior, while SNAP and FS-2 produce relatively smaller vacancy clusters. Overall, the maximum cluster size for both interstitials and vacancies increases with PKA energy for all potentials, although the exact values and trends differ based on the potential used.

\subsection{Defect Cluster Morphology Classification}

To better understand the types of defects created during displacement cascades, we classified the resulting clusters into six main morphological categories using a graph-based method~\cite{BHARDWAJ2021110474}. These include dislocation loops of type $\frac{1}{2}\langle 111 \rangle$,  $\langle 110 \rangle$, and  $\langle 100 \rangle$, as well as compact C15-like ring structures, disordered crowdion arrangements, and hybrid dislocation–ring forms. For clarity in the figures, each type is represented by a simple ASCII symbol: \texttt{||-111} for $\frac{1}{2}\langle 111 \rangle$ loops, \texttt{||-110} for  $\langle 110 \rangle$ loops, \texttt{||-100} for  $\langle 100 \rangle$ loops, \texttt{@} for C15-like rings, \texttt{\#} for disordered clusters, and \texttt{||@}, \texttt{||//} for hybrid structures.\\

As seen in Figure~\ref{fig:cluster_fraction}, the relative distribution of cluster types differs strongly between the four interatomic potentials—EAM, FS-1, SNAP, and FS-2. Most defects are dominated either by C15-like rings or by $\frac{1}{2}\langle 111 \rangle$  dislocation loops. FS-2 primarily produces $\frac{1}{2}\langle 111 \rangle$ loops, which account for more than 70\% of its defects, highlighting a strong tendency toward loop-dominated damage. It also generates some C15-like rings and disordered clusters, with only a few rare $\langle 110 \rangle$ dumbbells. In comparison, FS-1 and SNAP yield a larger fraction of C15-like rings, indicating more compact and localized defects with limited long-range dislocation growth. FS-1 additionally produces a noticeable share of $\langle 110 \rangle$ dumbbells, along with small loops and very few $\langle 111 \rangle$ dumbbells, while SNAP forms small percentage of both $\langle 111 \rangle$ and $\langle 110 \rangle$ dumbbells. EAM shows a more even distribution, forming both C15-like rings (about 43\%) and $\frac{1}{2}\langle 111 \rangle$ loops (about 39\%). It also produces a noticeable fraction of $\langle 110 \rangle$ clusters. Figure~\ref{fig:cluster_size_morph} shows how cluster sizes vary with morphology and potential. Dislocation loops, especially $\frac{1}{2}\langle 111 \rangle$, as well as hybrid and mixed morphologies, consistently form the largest clusters in all models. FS-2 generates the largest $\frac{1}{2}\langle 111 \rangle$ loops, often above 30 defects. FS-1 and SNAP also produce large hybrid dislocation–ring clusters, indicating that these potentials favor more complex dislocation networks. In contrast, C15-like rings and disordered clusters remain relatively small, typically between 3 and 9 defects. These are most common in FS-1 and SNAP, suggesting that these potentials favor compact, localized defects with limited growth after cascade events. The mixed dislocation–ring clusters are less frequent but usually large, particularly in FS-1 and FS-2, where they exceed 30 defects and may represent intermediate stages in defect evolution. Overall, the results indicate that FS-2 strongly favors the growth of large, loop-dominated clusters; FS-1 and SNAP are characterized by more compact C15-like defects of limited size; and EAM represents an intermediate case with a balanced mix of morphologies. These observations highlight that the choice of interatomic potential significantly influences not only the type of defect that forms but also its size and growth behavior.

\section{Discussion}
Previous studies on tungsten and iron \cite{BYGGMASTAR2018530,SAND2016119,TERENTYEV200665} have shown that the stiffness-to-range ratio (|S/R|) strongly influences cascade morphology. A higher |S/R| produces cascades that are more spread out and less dense due to extended ballistic energy transfer to surrounding atoms, whereas a lower |S/R| results in denser cascades with stronger local energy transfer \cite{BYGGMASTAR2018530,SAND2016119,TERENTYEV200665}. Consequently, higher |S/R| values generally lead to fewer peak defects, while lower |S/R| values produce higher peak defect numbers \cite{BYGGMASTAR2018530,SAND2016119,TERENTYEV200665}. Moreover, because cascades with higher |S/R| are more spread out, in-cascade recombination is less effective, resulting in a larger fraction of surviving defects. In contrast, the higher density cascades generated at lower |S/R| promote more recombination and reduce surviving defects \cite{SAND2016119}. In iron \cite{BYGGMASTAR2018530}, vacancy clustering has also been reported to depend on |S/R|, with higher |S/R| values favoring larger vacancy clusters. We discuss our results in the backdrop of these trends. Table~\ref{tab:stiffness_comparison} shows that EAM has the highest |S/R|, SNAP the lowest, while FS-1 and FS-2 lie in between. Figure~\ref{fig:interstitial_clusters} reveals that interstitial clustering is highest for SNAP, indicating denser cascades where recombination is more efficient, followed by FS-2, then EAM, and lowest for FS-1, where the cascades are more spread out and of lower density. Despite having $|S/R|$ values similar to FS-2, FS-1 shows a clustering behavior that does not follow the expected pattern.\\

From Figure~\ref{fig:potential_compare_a}, at 10 keV PKA, the peak number of defects is highest for EAM, slightly lower for FS-1, followed by FS-2, and lowest for SNAP. This trend does not directly follow |S/R| expectations, since one would anticipate SNAP to show the highest defect peaks and EAM the lowest. The discrepancy arises because, as shown in Table~\ref{tab:Nb_properties}, SNAP has the highest average threshold displacement energy (TDE) and defect formation energies, which suppress defect production. Conversely, EAM has lower defect formation energies and a lower average TDE, resulting in a larger peak number of defects. Although FS-1 and FS-2 have similar |S/R| values, FS-1 exhibits more defects due to its lower average TDE and defect formation energies compared to FS-2. At 50 keV (Figure~\ref{fig:potential_compare_b}), the peak defect numbers for EAM, FS-2, and SNAP converge, while FS-1 produces the highest number of peak defects. This cannot be fully explained by |S/R|, TDE, or defect formation energies. Instead, the formation of subcascades (Figure~\ref{fig:subcascades} at higher PKA energies (above the $\sim$20 keV threshold) likely governs the evolution of cascades. This is also consistent with Figure~\ref{fig:powerlaw_plot}, where the surviving defects follow a power-law distribution with two distinct slopes around the subcascade threshold. Vacancy clustering trends provide further insights. Figures~\ref{fig:vacancy_clusters} and \ref{fig:max_vac_cluster} show that EAM consistently produces the highest vacancy clustering fraction and the largest vacancy cluster sizes across all energies, consistent with earlier observations \cite{BYGGMASTAR2018530} that higher |S/R| promotes larger vacancy clusters, despite EAM also having the highest vacancy formation energy. Overall, the contradictory behavior of FS-1 in both interstitial clustering and peak defect production at higher energies suggests that cascade evolution cannot be explained solely by |S/R| or TDE. Instead, it is governed by the combined influence of multiple material- and potential-dependent properties, which collectively shape defect production, recombination, and clustering.\\  

Vacancy clustering always exceeds interstitial clustering across all potentials (Figure~\ref{fig:interstitial_clusters} vs. Figure~\ref{fig:vacancy_clusters}). This differs from the findings of Lin et al.~(\cite{lin2017molecular}) in niobium, where vacancy and interstitial clustering were nearly the same, both around $\sim20\%$ beyond 5~keV up to 20~keV. In our study, the percentage of interstitial clustering stays above $30\%$ (up to 75~keV) for all four potentials, while vacancy clustering beyond 5~keV remains above $60\%$ (up to 75~keV). Our results also suggest that most defects exist as single interstitials, since interstitial clustering is low (Figure~\ref{fig:interstitial_clusters}) and the maximum cluster size is small (Figure~\ref{fig:max_inter_cluster}). This behavior is unlike tungsten \cite{Bhardwaj_2021}, where interstitial clustering is more pronounced. Thus, while tungsten shows stronger interstitial cluster formation, niobium exhibit a clear dominance of vacancy clustering over interstitial clustering. However, our results align with Qiu et al. \cite{QIU2023101394} in vanadium, suggesting that stronger vacancy clustering relative to interstitial clustering may be a characteristic of certain refractory BCC metals under irradiation. The trends in vacancy cluster size (Figure~\ref{fig:vacancy_clusters}) and clustering fraction (Figure~\ref{fig:vacancy_clusters}) indicate that early-stage vacancy aggregation occurs for all potentials, providing precursors for void nucleation. At higher temperatures and irradiation doses (dpa), these clusters may further evolve into voids or self-organize into ordered void superlattices. Such evolution is consistent with experimental observations \cite{Brimhall01011973}, where niobium developed void structures and void lattices under high-dose irradiation.\\

Simulation results indicate that the morphology of defect clusters in niobium is highly sensitive to the choice of interatomic potential. Among the models tested, FS-2 notably generates the highest number of extended 1/2$\langle$111$\rangle$ dislocation loops, consistent with previous molecular dynamics studies on bcc metals such as iron, molybdenum, tungsten, and vanadium~\cite{Stoller2012_PrimaryDamage,Pasianot01062002,QIU2023101394,Bhardwaj_2021}, as well as experimental observations in niobium~\cite{Tucker01091967}. FS-1 and SNAP tend to generate a larger fraction of C15-like rings, disordered clusters, and hybrid morphologies (Figure~\ref{fig:cluster_fraction}, Figure~\ref{fig:cluster_size_morph}), which may be related to their training on high-temperature or liquid-phase data. These potentials also show a higher fraction of C15-like ring defects, which have been reported previously in tungsten and are recognized as a common damage structure in bcc metals under irradiation~\cite{Bhardwaj_2021,warrier2024defectclustermorphologyw}. Similar behavior has been observed in iron, where molecular dynamics simulations revealed that C15 clusters formed in cascades can remain stable or collapse into dislocation loops. When C15 clusters grow, they can eventually collapse, most often transforming into $1/2\langle111\rangle$ loops and occasionally into $\langle100\rangle$ loops, while leaving behind small stable C15 rings~\cite{BYGGMASTAR2020151893}. This suggests that the C15 rings observed in Nb may also serve as precursors that evolve into $\langle111\rangle$ or $\langle100\rangle$ loops, further supporting their importance in cascade damage processes of bcc metals. The EAM potential generated a mix of loop- and ring-like clusters. $\langle 100 \rangle$ clusters were mostly absent, with only rare instances appearing, indicating that they are likely short-lived or unstable. A comparable behavior has been observed in tungsten, where molecular dynamics studies show that $\langle 100 \rangle$ loops are less stable than $\tfrac{1}{2}\langle 111 \rangle$ loops and usually evolve into more stable forms unless they reach sufficiently large sizes \cite{Bhardwaj2022}.

All four potentials are capable of producing $\tfrac{1}{2}\langle 111 \rangle$, 
$\langle 110 \rangle$ loops, and C15 ring defects, but FS-2 reproduces the dominant experimental defect types most accurately. At higher energies, or under successive collision cascades, the C15 ring clusters can grow larger across all potentials. In particular, SNAP tends to form larger C15 ring clusters, which may subsequently transform into $\tfrac{1}{2}\langle 111 \rangle$ loops, $\langle 100 \rangle$ loops, and smaller stable C15 clusters. The rare $\langle 100 \rangle$ features observed with the EAM potential at high energies suggest that similar configurations may also emerge in other potentials at even higher energies or during overlapping cascades. Therefore, additional simulations considering successive or overlapping cascades are recommended to better understand their formation mechanisms.

\section{Conclusion}

This study employs molecular dynamics simulations to examine radiation damage in pure niobium across a broad range of primary knock-on atom (PKA) energies (1--75 keV) at 300 K. Four interatomic potentials were compared: EAM, FS-1, FS-2, and SNAP. While all potentials captured the general features of cascade formation, they exhibited clear differences in defect production, clustering behavior, and damage morphology. At low energies, prior to subcascade formation, defect production, recombination, and clustering generally followed the expected trends governed by the combined effects of the stiffness-to-range ratio (|S/R|), threshold displacement energy (TDE), and defect formation energies, with the notable exception of FS-1. At higher energies, once subcascades emerged, these parameters alone were insufficient to explain the observed behavior, as cascade evolution became more complex. Defect morphologies were found to be highly sensitive to the interatomic potential. The EAM potential generated a balanced mix of loop- and ring-like clusters and uniquely predicted occasional $\langle 100 \rangle$ cluster at high energies. FS-1 predominantly produced $\langle 110 \rangle$ dumbbells, crowdions, and small C15 rings, along with some $\tfrac{1}{2}\langle 111 \rangle$ crowdions and hybrid or disordered structures. FS-2 primarily formed $\tfrac{1}{2}\langle 111 \rangle$ dumbbells and crowdions, and at higher energies produced $\tfrac{1}{2}\langle 111 \rangle$ dislocation loops, consistent with behavior in other bcc metals. Smaller fractions of $\langle 110 \rangle$ dumbbells, C15 rings, and hybrid morphologies were also observed. The SNAP potential, in contrast, produced compact cascades with some disordered clusters, a significant fraction of $\langle 111 \rangle$ dumbbells, and a relatively high proportion of C15-like rings, with sizes reaching up to 9. These rings may act as precursors to $\tfrac{1}{2}\langle 111 \rangle$ and $\langle 100 \rangle$ dislocation loops as they evolve.\\

Overall, these findings give important insights into how niobium responds to irradiation and the types of microstructural changes that can develop under energetic particle exposure. Our study highlights a clear difference between older classical potentials and the newer ones, with the older potentials producing more defects and giving the wrong ground-state configuration of SIAs compared to the newer potentials.

\section*{Acknowledgements}

We gratefully acknowledge the CAD HPC team for their support in maintaining the computing resources and assisting with the installation of LAMMPS.

\section*{Disclosure statement}
No competing interests are associated with this work.

\section*{Funding}
This work was carried out without financial support from any funding agency.

\printbibliography

\newpage
\section*{Tables}

\begin{table}[H]
\centering
\caption{Elastic, thermal, and defect properties of Nb. 
Values in parentheses show the relative error (\%) with respect to DFT or experiment. 
Symbols: $a_{0}$ = lattice parameter, $C_{11}, C_{12}, C_{44}$ = elastic constants, 
$B$ = bulk modulus, $G$ = shear modulus, $\nu$ = Poisson’s ratio, 
$T_{m}$ = melting point, $\alpha$ = thermal expansion coefficient, 
$E_{f}^{v}$ = vacancy formation energy, $E_{m}^{v}$ = vacancy migration energy, 
Tet = tetrahedral SIA, Oct = octahedral SIA, $\langle hkl \rangle$ = dumbbell SIA orientations, 
$E_{d}^{avg}$ = average threshold displacement energy. 
The DFT data for lattice parameters, elastic constants, and defect formation and migration energies are taken from Refs.~\cite{r11,r12,r13}, while melting point values are from Refs.~\cite{r56,r57}. 
Experimental values for lattice parameters, elastic constants, thermal expansion, and melting point are from Refs.~\cite{r53,r54}, and data for vacancy formation, migration energies, and threshold displacement energies are from Refs.~\cite{r62,r63}. 
For reference, the ASTM standard threshold displacement energy of 60~eV is also widely used in the literature~\cite{r64,r65}, although it is not included in Table~\ref{tab:Nb_properties}.}
\scriptsize
\setlength{\tabcolsep}{4pt}
\renewcommand{\arraystretch}{1.2}
\begin{tabular}{lccccc}
\hline
\textbf{Property} & \textbf{SNAP} & \textbf{FS-2} & \textbf{FS-1} & \textbf{EAM} & \textbf{DFT/Exp.} \\
\hline
\multicolumn{6}{c}{\textbf{Elastic \& Thermal Properties}} \\
$a_{0}$ (\AA) & 3.31 (0.0\%) & 3.312 (0.1\%) & 3.327 (0.5\%) & 3.314 (0.1\%) & 3.31 / 3.30 \\
$C_{11}$ (GPa) & 236 (–0.4\%) & 238 (0.4\%) & 202 (\textbf{–14.8\%}) & 225 (–5.1\%) & 237 / 247 \\
$C_{12}$ (GPa) & 135 (0.0\%) & 130 (–3.0\%) & 78 (\textbf{–41.5\%}) & 127 (–5.2\%) & 134 / 135 \\
$C_{44}$ (GPa) & 14 (0.0\%) & 17 (17.9\%) & 20 (\textbf{57.1\%}) & 14 (0.0\%) & 14 / 28 \\
$B$ (GPa) & 169 (–0.2\%) & 166 (–1.8\%) & 120 (\textbf{–29.0\%}) & 160 (–5.3\%) & 169 / 170.3 \\
$G$ (GPa) & 29 (0.0\%) & 42 (\textbf{40.0\%}) & 51 (\textbf{70.0\%}) & 39 (\textbf{30.0\%}) & 29 / 30 \\
$\nu$ & 0.43 (7.5\%) & 0.35 (–7.9\%) & 0.28 (\textbf{–26.3\%}) & 0.36 (–5.3\%) & 0.39 / 0.38--0.40 \\
$T_{m}$ (K) & 2800 (1.8\%) & 3000 (9.1\%) & 3050 (10.9\%) & 3000 (9.1\%) & 2750 / 2750 \\
$\alpha$ ($10^{-6}$ K$^{-1}$) & 6.7 (–8.2\%) & 10.7 (\textbf{33.8\%}) & 7.5 (–6.3\%) & 9.7 (\textbf{21.3\%}) & 8.0 / 7.3 \\
\multicolumn{6}{c}{\textbf{Defect Formation \& Migration Energies}} \\
$E_{f}^{v}$ (eV) & 2.80 (1.1\%) & 2.83 (1.4\%) & 2.81 (1.4\%) & 3.13 (4.3\%) & 2.77 / 2.60--3.00 \\
$E_{m}^{v}$ (eV) & 0.68 (4.6\%) & 0.93 (\textbf{26.2\%}) & 0.82 (\textbf{20.0\%}) & 0.78 (\textbf{20.0\%}) & 0.65 / 0.55 \\
Tet (eV) & 4.45 (0.7\%) & 4.60 (4.1\%) & 3.84 (–13.1\%) & 3.92 (–11.3\%) & 4.42 / --- \\
Oct (eV) & 4.77 (3.2\%) & 4.76 (2.8\%) & 4.10 (–11.3\%) & 4.72 (2.2\%) & 4.62 / --- \\
$\langle 111 \rangle$ (eV) & 4.08 (3.3\%) & 4.01 (1.5\%) & 3.73 (–5.6\%) & 4.02 (1.8\%) & 3.95 / --- \\
$\langle 110 \rangle$ (eV) & 4.25 (1.2\%) & 4.20 (0.0\%) & 3.77 (–10.3\%) & 3.77 (–10.3\%) & 4.20 / --- \\
$\langle 100 \rangle$ (eV) & 4.60 (2.2\%) & 4.63 (2.9\%) & 4.19 (–6.9\%) & 4.45 (–1.1\%) & 4.50 / --- \\
$E_d^{\text{avg}}$ (eV) & 77.7 (–5.8\%) & 56.9 (\textbf{–31.0\%}) & 40.0 (\textbf{–51.5\%}) & 51.0 (\textbf{–38.2\%}) & --- / 78--87$^{a}$ \\
\hline
\end{tabular}

\begin{flushleft}
\footnotesize $^{a}$ For $E_d^{\mathrm{avg}}$, deviations are calculated relative to 82.5 eV (the midpoint of the experimental range 78--87 eV).
\end{flushleft}
\label{tab:Nb_properties}
\end{table}

\newpage
\begin{table}[H]
    \centering
    \caption{Comparison of short-range interaction parameters for different interatomic potentials.}
    \label{tab:stiffness_comparison}
    \begin{tabular}{lcccc}
        \hline
        Parameter & SNAP & EAM & FS-1 & FS-2 \\
        \hline
        \( R \) (\AA) & 1.34 & 1.12 & 1.23 & 1.34 \\
        \( S \) (eV/\AA) & -102.35 & -239.35 & -132.98 & -152.12 \\
        \( |S/R| \) (eV/\AA$^2$) & 76.13 & 213.25 & 107.36 & 113.39 \\
        \hline
    \end{tabular}
\end{table}

\newpage
\begin{table}[h!]
\centering
\caption{Simulation box sizes and number of atoms for different interatomic potentials and PKA energies.}
\scriptsize   
\begin{tabular}{c c c c c}
\hline
\textbf{PKA Energy (keV)} & \multicolumn{2}{c}{\textbf{SNAP / FS-2}} & \multicolumn{2}{c}{\textbf{EAM / FS-1}} \\
\cline{2-3}\cline{4-5}
& Box size & Atoms & Box size & Atoms \\
\hline
1   & $40 \times 40 \times 40$   & 128,000     & $40 \times 40 \times 40$   & 128,000 \\
5   & $80 \times 80 \times 80$   & 1,024,000   & $80 \times 80 \times 80$   & 1,024,000 \\
10  & $120 \times 120 \times 120$& 3,456,000   & $120 \times 120 \times 120$& 3,456,000 \\
20  & $180 \times 180 \times 180$& 11,664,000  & $180 \times 180 \times 180$& 11,664,000 \\
30  & $210 \times 210 \times 210$& 18,522,000  & $220 \times 220 \times 220$& 21,296,000 \\
50  & $260 \times 260 \times 260$& 35,152,000  & $280 \times 280 \times 280$& 43,904,000 \\
75  & $320 \times 320 \times 320$& 65,536,000  & $360 \times 360 \times 360$& 93,312,000 \\
\hline
\end{tabular}
\label{tab:box}
\end{table}

\newpage
\begin{table}[H]
    \centering
    \caption{Fitted parameters for the power-law relation \(N_{\text{FP}} = a \cdot E^b\) in two energy regimes for different interatomic potentials in niobium. Here, \(a_1, b_1\) correspond to region~1 (1--20~keV), and \(a_2, b_2\) correspond to region~2 (20--75~keV).}
    \label{tab:powerlaw_fit}
    \begin{tabular}{lcccc}
        \hline
        Potential & \(a_1\) & \(b_1\) & \(a_2\) & \(b_2\) \\
        \hline
        EAM   & 7.539 & 0.683 & 5.010 & 0.841 \\
        FS-1  & 5.882 & 0.747 & 3.254 & 0.930 \\
        SNAP  & 2.040 & 0.864 & 1.802 & 0.902 \\
        FS-2  & 2.697 & 0.757 & 1.770 & 0.902 \\
        \hline
    \end{tabular}
\end{table}

\newpage
\section*{Figures}

\begin{figure}[H]
    \centering
    \includegraphics[width=0.9\textwidth]{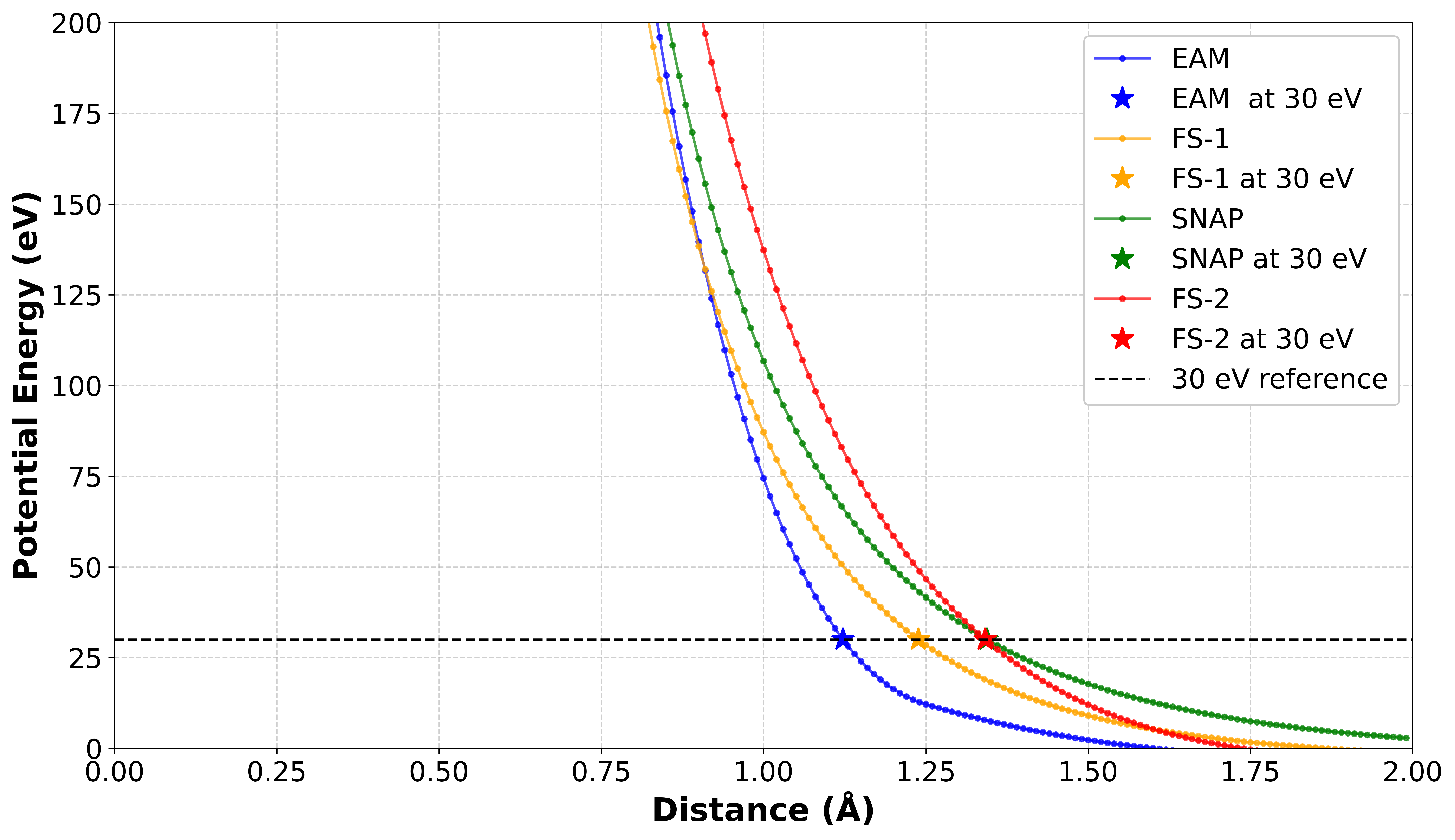}
    \caption{Pairwise potential energy curves at short interatomic distances for
different interatomic potentials in niobium (SNAP, EAM, FS-1, Fs-2). Stars indicate the distance R
where the potential energy reaches 30~eV.}
    \label{fig:short_range}
\end{figure}

\begin{figure}[H]
    \centering

    \begin{subfigure}{0.9\textwidth}
        \centering
        \includegraphics[width=\textwidth]{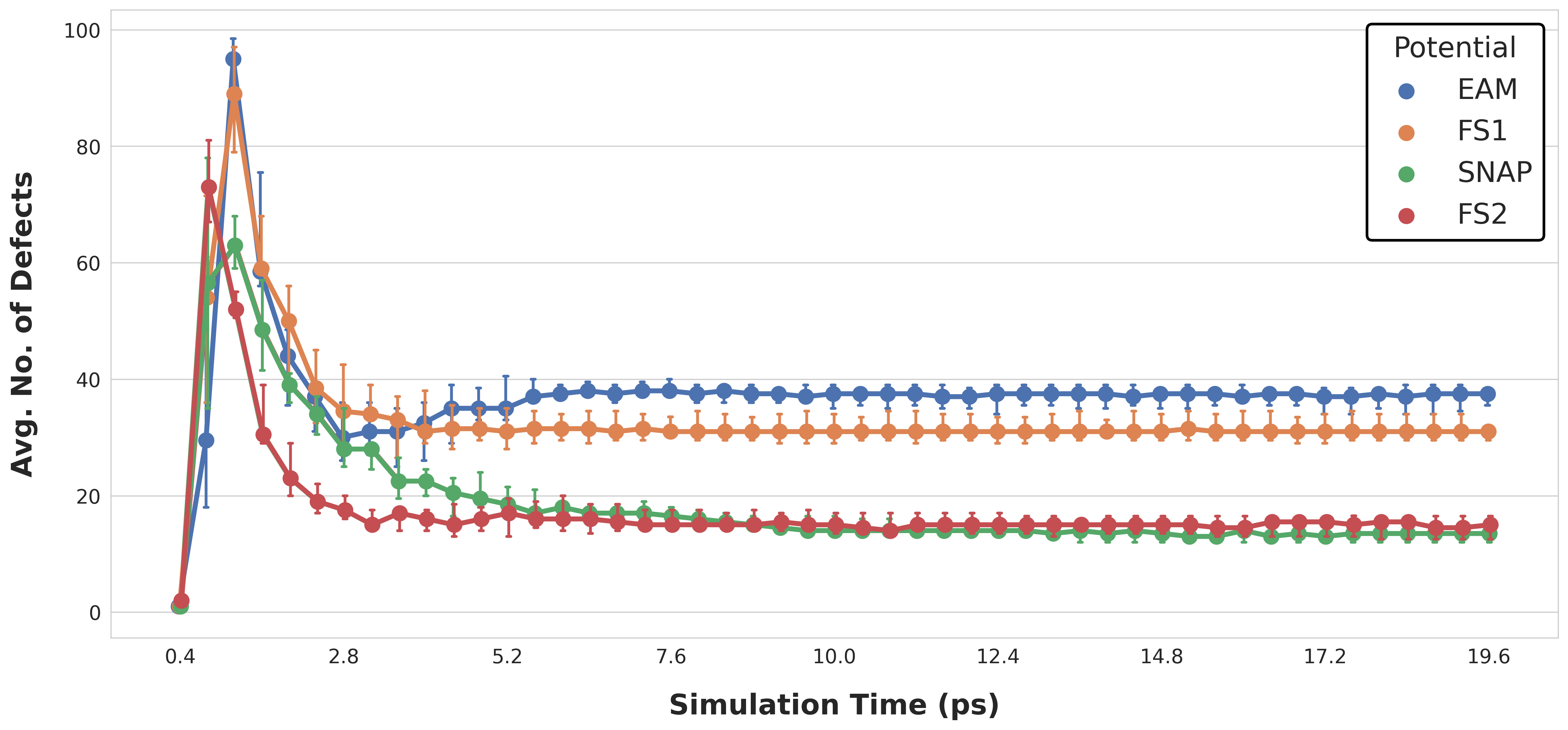}
        \caption{}
        \label{fig:potential_compare_a}
    \end{subfigure}

    \vspace{0.5cm}

    \begin{subfigure}{0.9\textwidth}
        \centering
        \includegraphics[width=\textwidth]{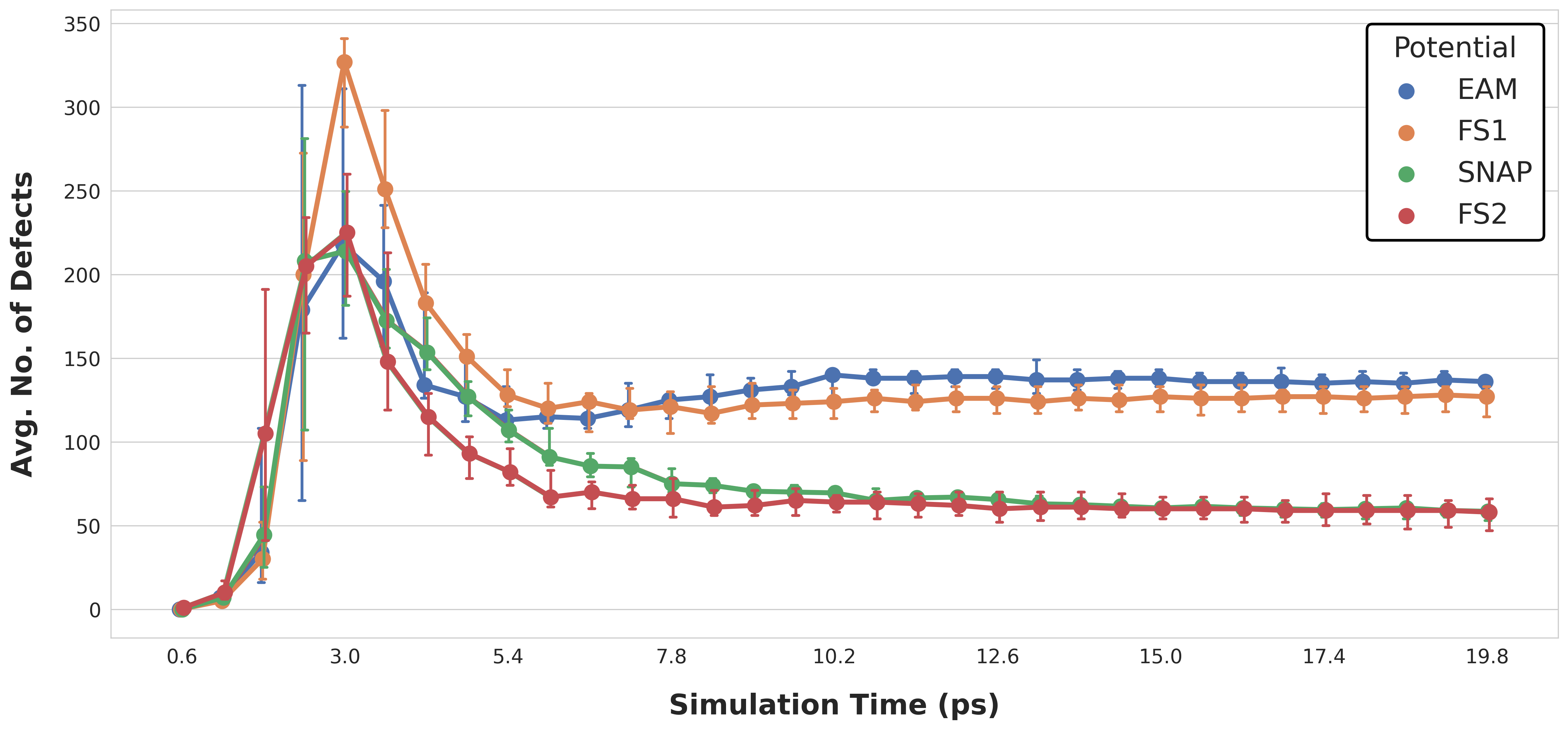}
        \caption{}
        \label{fig:potential_compare_b}
    \end{subfigure}

    \caption{Cascade evolution in pure Nb at two different PKA energies using four interatomic potentials (SNAP, EAM, FS-1, FS-2).  
    The subfigures show: (a) 10~keV PKA, and (b) 50~keV PKA. Error bars indicate statistical variation across 25 cascade simulations.}
    \label{fig:potential_compare}
\end{figure}

\begin{figure}[H]
    \centering
    \includegraphics[width=0.9\textwidth]{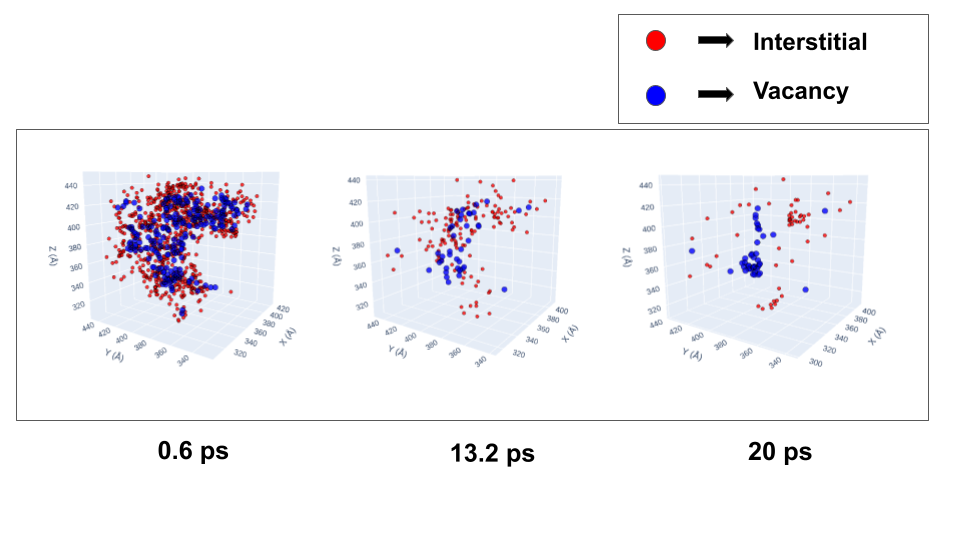}
    \caption{Snapshots showing the temporal evolution of defects produced by a 
\SI{50}{\kilo\electronvolt} PKA cascade in Nb using the SNAP potential, at three 
representative times: \SI{0.6}{\pico\second}, \SI{13.2}{\pico\second}, and 
\SI{20}{\pico\second}. Red spheres denote interstitials and blue spheres denote 
vacancies. While these snapshots are from a SNAP simulation, the general stages 
of defect production, recombination, and stabilization are common to all 
interatomic potentials.}

    \label{fig:defect_evolution}
\end{figure}

\begin{figure}[H]
    \centering
    \includegraphics[width=0.9\textwidth]{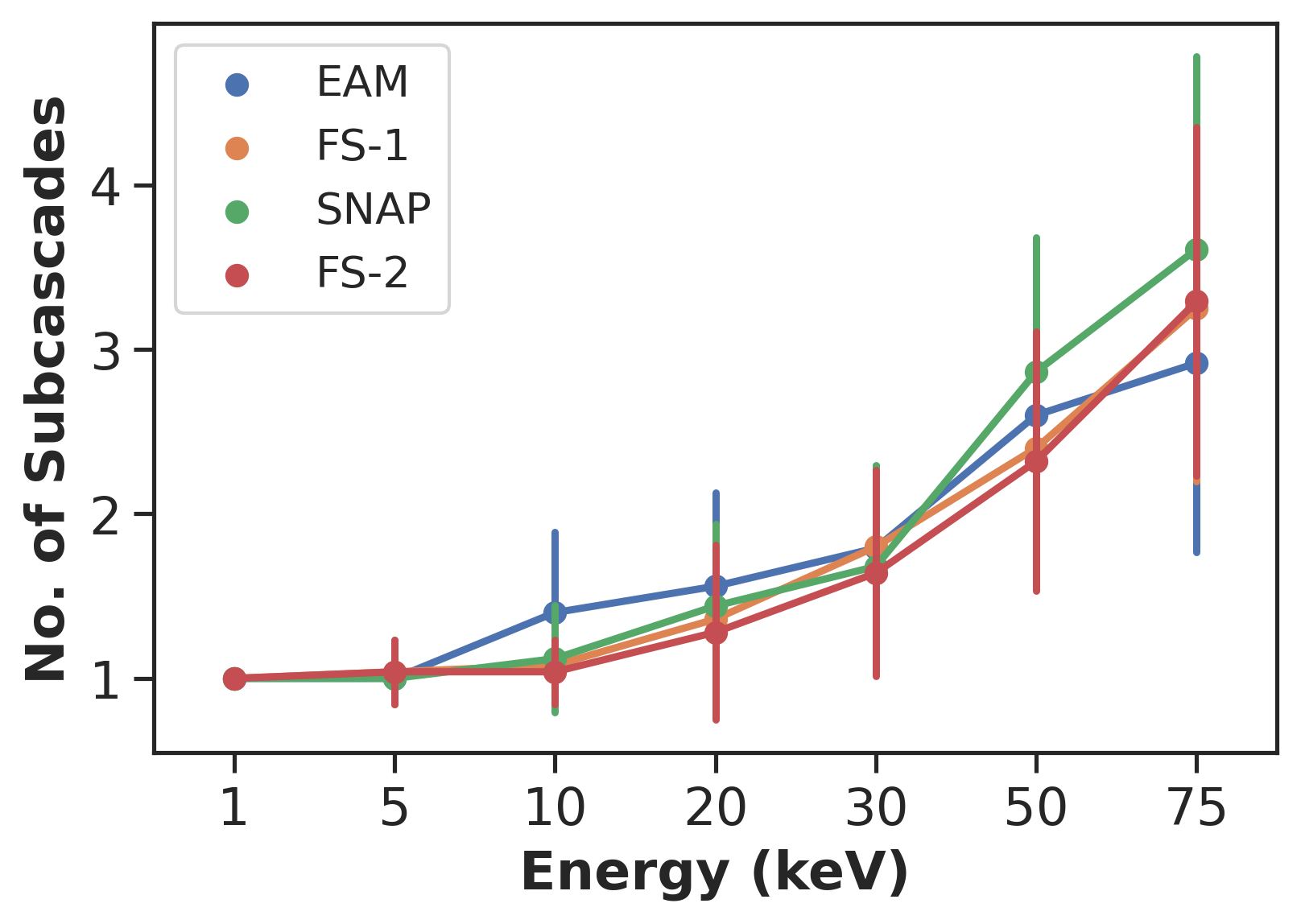}
    \caption{Average number of subcascades as a function of PKA energy for SNAP, EAM, FS-1, and FS-2 potentials. Subcascade formation starts around 20~keV.}
    \label{fig:subcascades}
\end{figure}

\begin{figure}[H]
    \centering
    \includegraphics[width=0.9\textwidth]{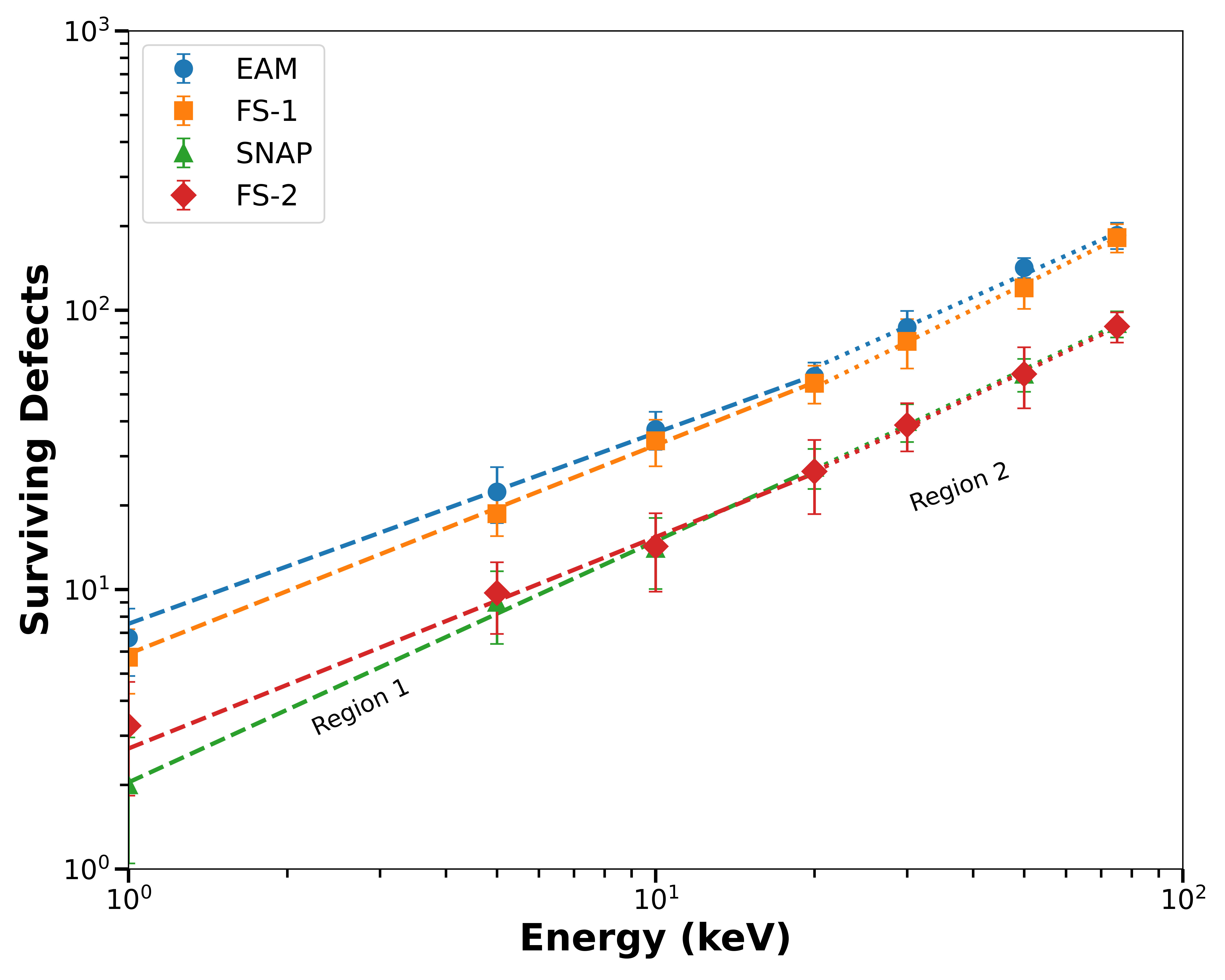} 
    \caption{Number of surviving defects as a function of PKA energy in niobium, calculated using four interatomic potentials: SNAP, EAM, FS-1, and FS-2.}

    \label{fig:powerlaw_plot}
\end{figure}

\begin{figure}[H]
\centering
\includegraphics[width=0.9\textwidth]{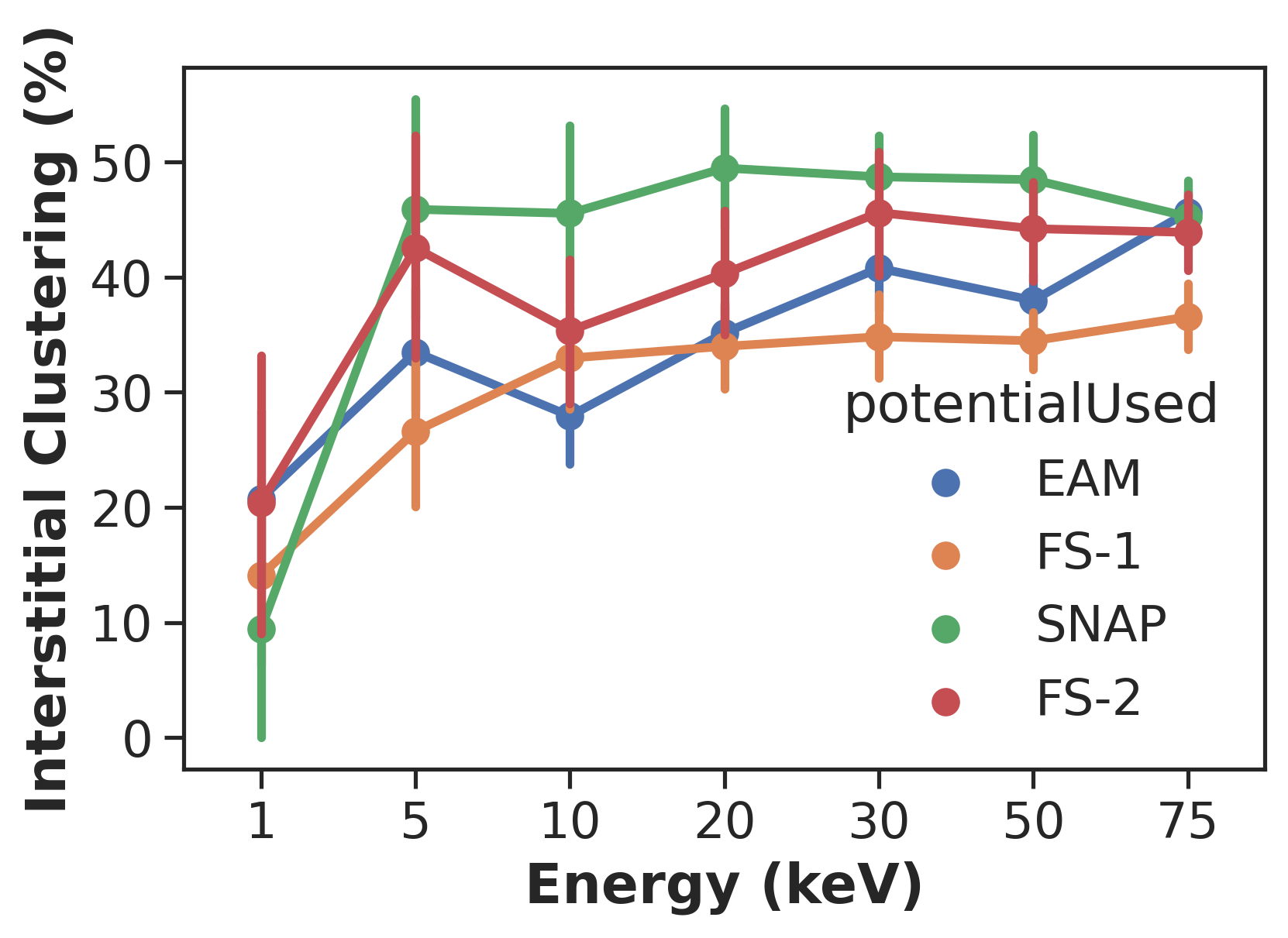}
\caption{Percentage of interstitials found in clusters as a function of PKA energy for different interatomic potentials (SNAP, EAM, FS-1, FS-2). Error bars represent variability across 25 cascade simulations.}
\label{fig:interstitial_clusters}
\end{figure}

\begin{figure}[H]
\centering
\includegraphics[width=0.9\textwidth]{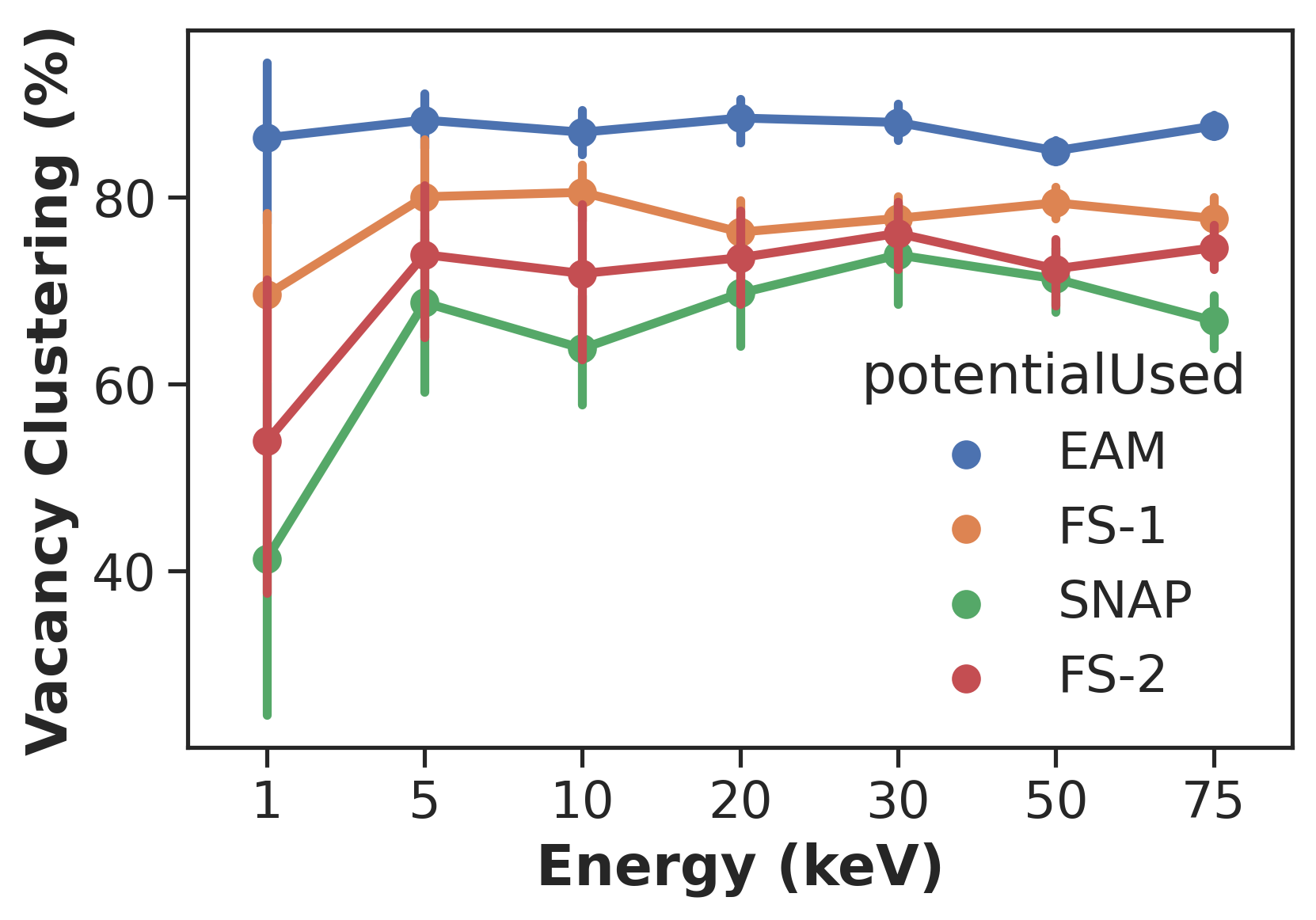}
\caption{Percentage of vacancies found in clusters as a function of PKA energy for different interatomic potentials (SNAP, EAM, FS-1, FS-2). Error bars indicate variation across simulation runs.}
\label{fig:vacancy_clusters}
\end{figure}

\begin{figure}[H]
    \centering
    \includegraphics[width=0.9\textwidth]{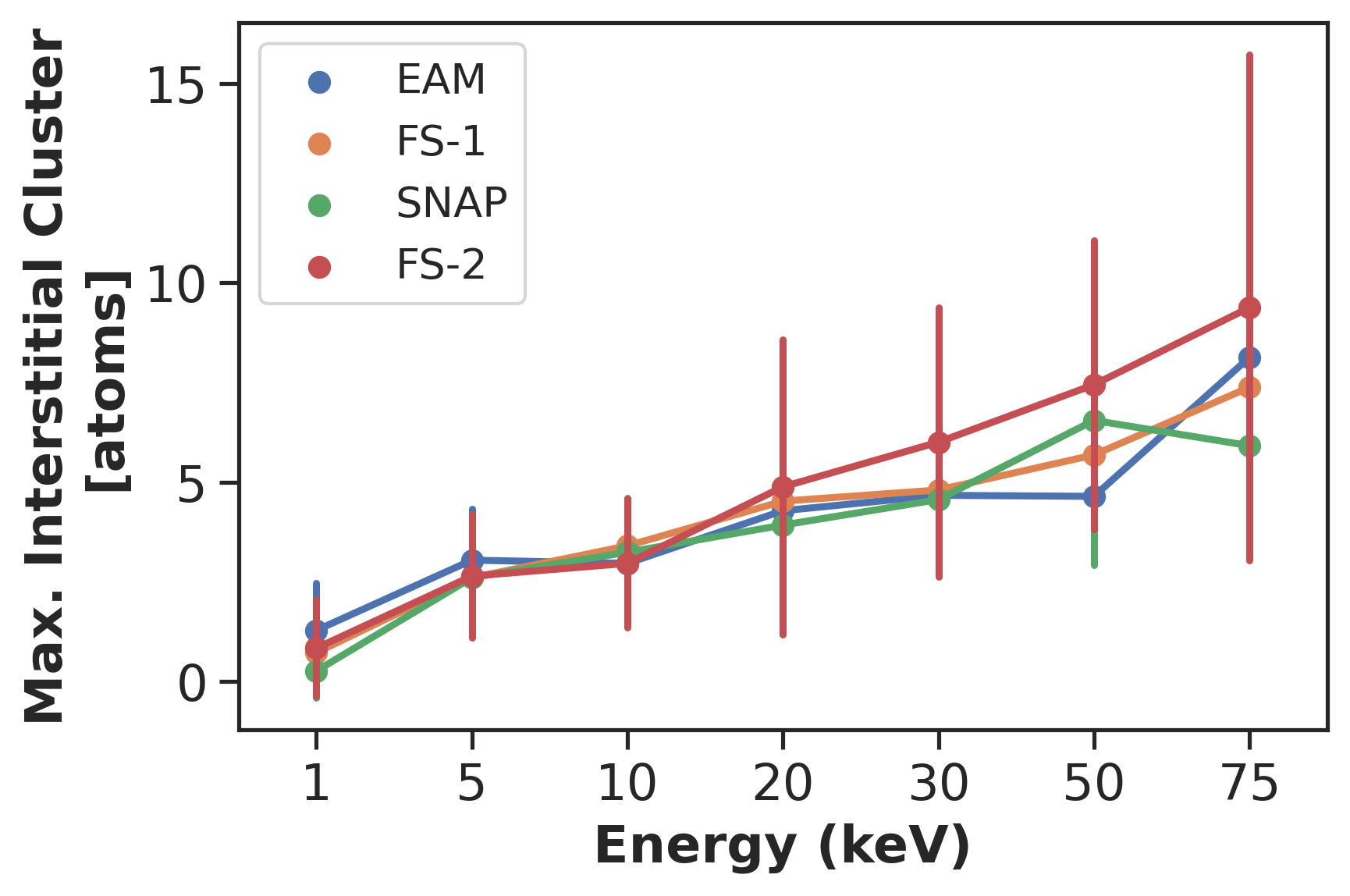}
    \caption{Maximum interstitial cluster size as a function of PKA energy using four interatomic potentials (SNAP, EAM, FS-1, FS-2).}
    \label{fig:max_inter_cluster}
\end{figure}

\begin{figure}[H]
    \centering
    \includegraphics[width=0.9\textwidth]{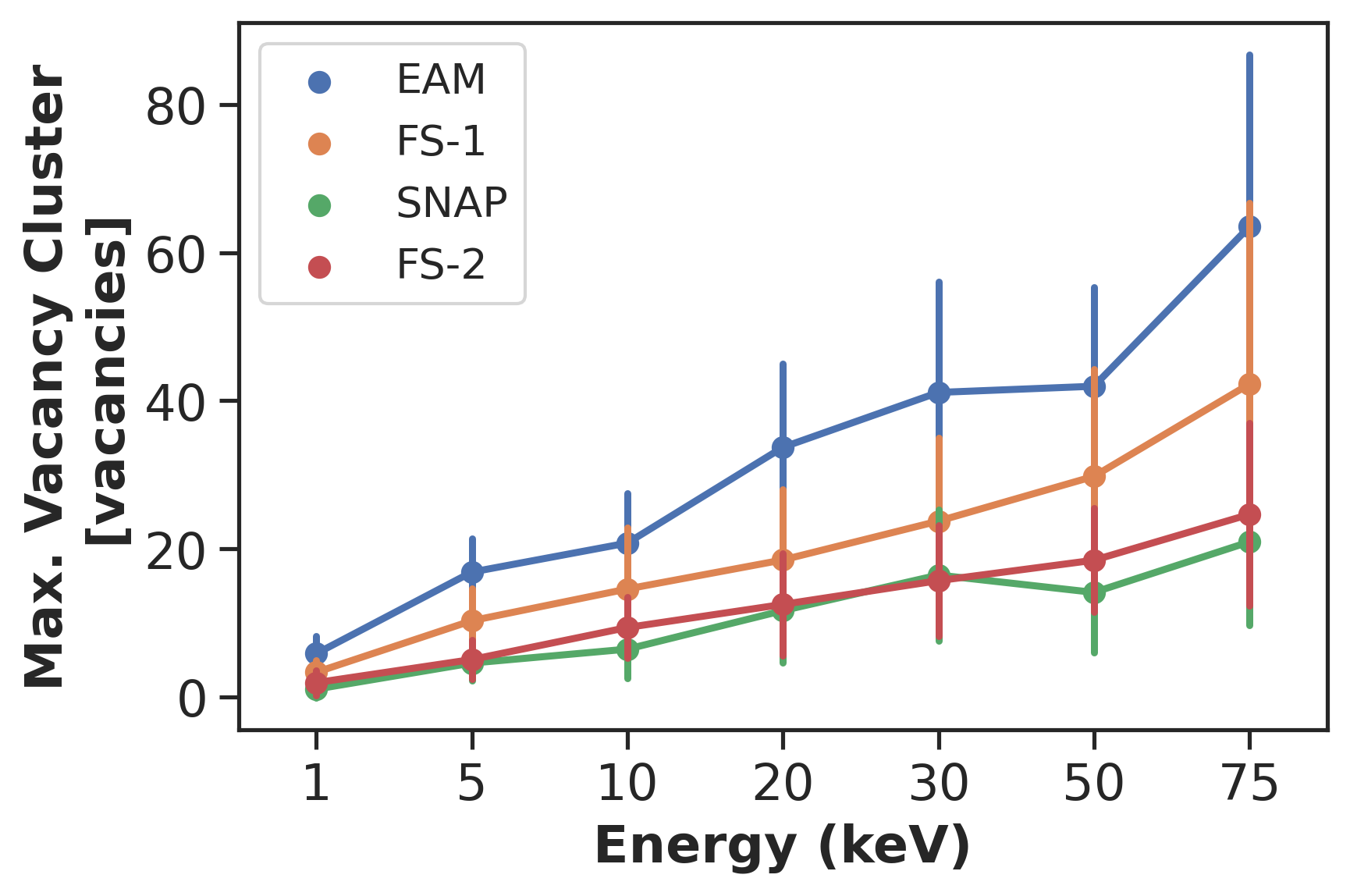}
    \caption{Maximum vacancy cluster size as a function of PKA energy using four interatomic potentials (SNAP, EAM, FS-1, FS-2).}
    \label{fig:max_vac_cluster}
\end{figure}

\begin{figure}[H]
    \centering
    \includegraphics[width=0.9\textwidth]{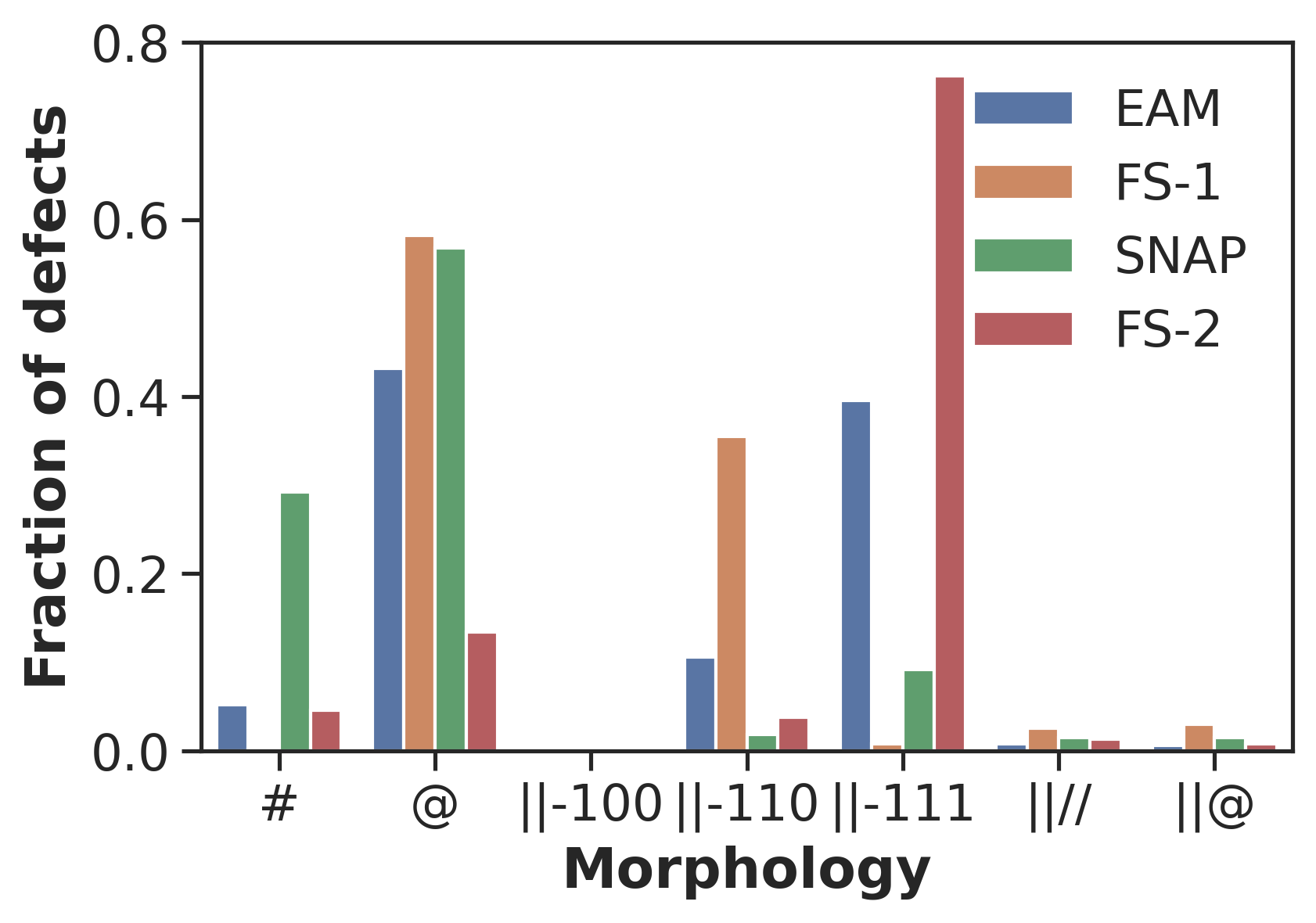}
    \caption{Fraction of clusters by morphology for SNAP, EAM, FS-1, FS-2 potentials.  
    Symbols: \texttt{\detokenize{||-111}} = $\frac{1}{2}\langle 111 \rangle$ loops, 
    \texttt{\detokenize{||-110}} = $\langle 110 \rangle$ loops, 
    \texttt{\detokenize{||-100}} = $\langle 100 \rangle$ loops, 
    \texttt{@} = C15-like rings, 
    \texttt{\#} = disordered clusters, 
    \texttt{\detokenize{||@}}, \texttt{\detokenize{||//}} = hybrid morphologies.}
    \label{fig:cluster_fraction}
\end{figure}

\begin{figure}[H]
    \centering
    \includegraphics[width=0.9\textwidth]{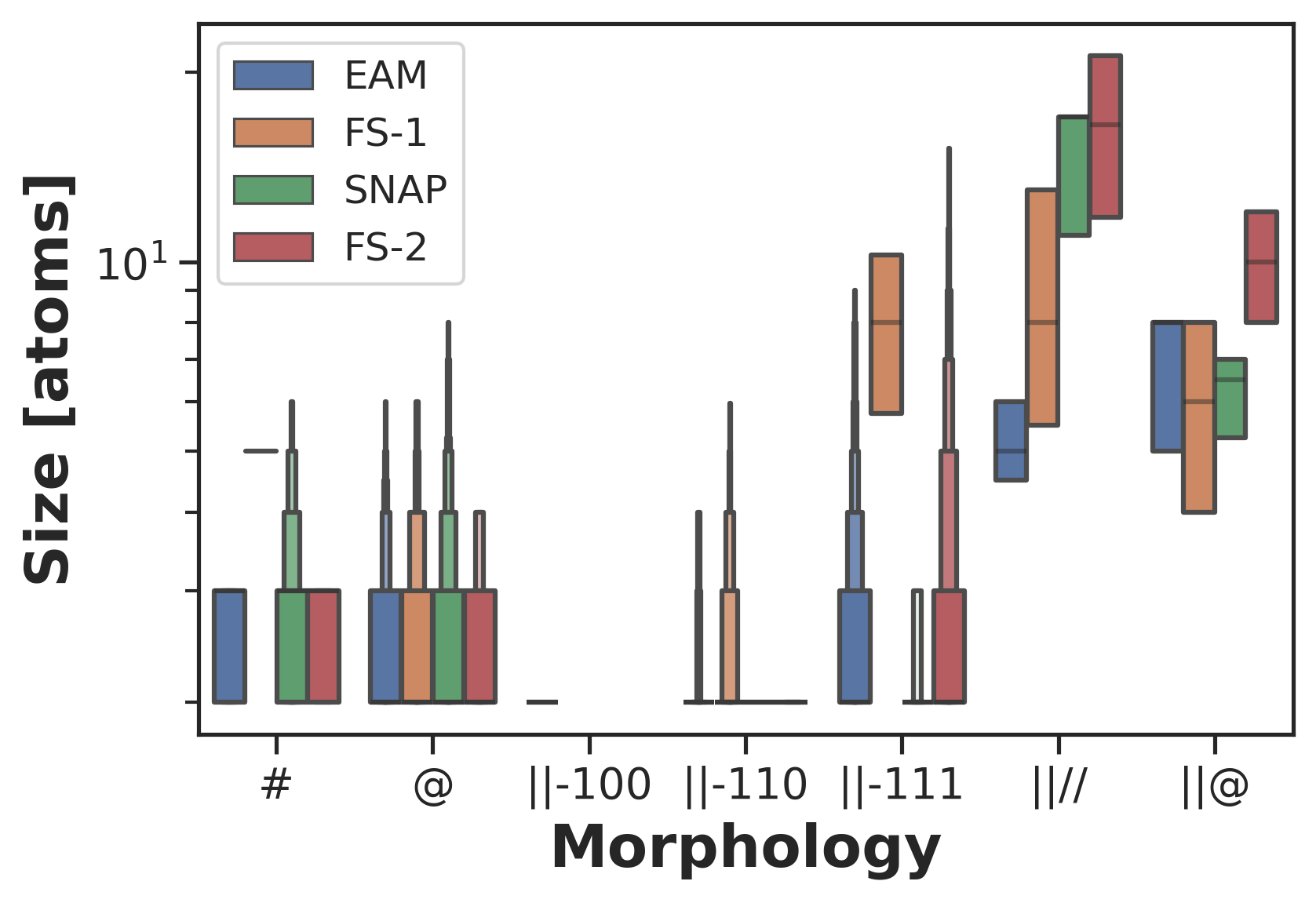}
    \caption{Cluster size distribution by morphology for four interatomic potentials (SNAP, EAM, FS-1, FS-2) in niobium, shown on a logarithmic scale. Symbols are the same as in Figure~\ref{fig:cluster_fraction}.}
    \label{fig:cluster_size_morph}
\end{figure}

\begin{figure}[H]
    \centering
    \includegraphics[width=0.9\textwidth]{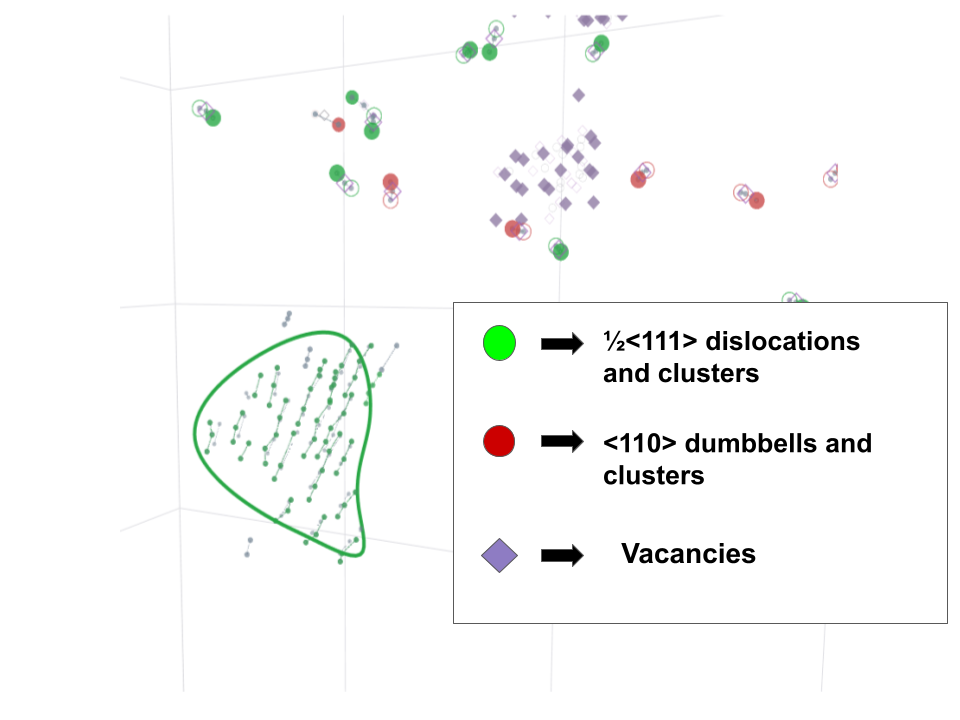}
    \caption{Atomistic snapshot of dislocation loop formation in pure niobium during a 75~keV PKA cascade simulated with the FS-2 potential. Green represent $\langle 111 \rangle$ dumbbells and dislocation loops, red represent $\langle 110 \rangle$ dumbbells, and violet indicate vacancies. The figure illustrates the formation of $\tfrac{1}{2}\langle 111 \rangle$ dislocation loops, consistent with the statistical trends observed in cascade simulations.}

    \label{fig:loop_morphology}
\end{figure}

\newpage
\section*{Figure Captions}
Figure~\ref{fig:short_range}. Pairwise potential energy curves at short interatomic distances for
different interatomic potentials in niobium (SNAP, EAM, FS-1, Fs-2). Stars indicate the distance R
where the potential energy reaches 30~eV.\\
\\
Figure~\ref{fig:potential_compare}. Cascade evolution in pure Nb at two different PKA energies using four interatomic potentials (SNAP, EAM, FS-1, FS-2).  
    The subfigures show: (a) 10~keV PKA, and (b) 50~keV PKA.\\
\\
Figure~\ref{fig:defect_evolution}. Snapshots showing the temporal evolution of defects produced by a 
\SI{50}{\kilo\electronvolt} PKA cascade in Nb using the SNAP potential, at three 
representative times: \SI{0.6}{\pico\second}, \SI{13.2}{\pico\second}, and 
\SI{20}{\pico\second}. Red spheres denote interstitials and blue spheres denote 
vacancies. While these snapshots are from a SNAP simulation, the general stages 
of defect production, recombination, and stabilization are common to all 
interatomic potentials.\\
\\
Figure~\ref{fig:subcascades}. Average number of subcascades as a function of PKA energy for SNAP, EAM, FS-1, and FS-2 potentials. Subcascade formation starts around 20~keV.\\
\\
Figure~\ref{fig:powerlaw_plot}. Number of surviving defects as a function of PKA energy in niobium, calculated using four interatomic potentials: SNAP, EAM, FS-1, and FS-2.\\
\\
Figure~\ref{fig:interstitial_clusters}. Percentage of interstitials found in clusters as a function of PKA energy for different interatomic potentials (SNAP, EAM, FS-1, FS-2). Error bars represent variability across 25 cascade simulations.\\
\\
Figure~\ref{fig:vacancy_clusters}. Percentage of vacancies found in clusters as a function of PKA energy for different interatomic potentials (SNAP, EAM, FS-1, FS-2). Error bars indicate variation across simulation runs. \\
\\
Figure~\ref{fig:max_inter_cluster}. Maximum interstitial cluster size as a function of PKA energy using four interatomic potentials (SNAP, EAM, FS-1, FS-2).\\
\\
Figure~\ref{fig:max_vac_cluster}. Maximum vacancy cluster size as a function of PKA energy using four interatomic potentials (SNAP, EAM, FS-1, FS-2).  \\
\\
Figure~\ref{fig:cluster_fraction}. Fraction of clusters by morphology for SNAP, EAM, FS-1, FS-2 potentials. 
    Symbols: \texttt{\detokenize{||-111}} = $\frac{1}{2}\langle 111 \rangle$ loops, 
    \texttt{\detokenize{||-110}} = $\langle 110 \rangle$ loops, 
    \texttt{\detokenize{||-100}} = $\langle 100 \rangle$ loops, 
    \texttt{@} = C15-like rings, 
    \texttt{\#} = disordered clusters, 
    \texttt{\detokenize{||@}}, \texttt{\detokenize{||//}} = hybrid morphologies.\\
\\
Figure~\ref{fig:cluster_size_morph}. Cluster size distribution by morphology for four interatomic potentials (SNAP, EAM, FS-1, FS-2) in niobium, shown on a logarithmic scale. Symbols are the same as in Figure~\ref{fig:cluster_fraction}.\\
\\
Figure~\ref{fig:loop_morphology}. Atomistic snapshot of dislocation loop formation in pure niobium during a 75~keV PKA cascade simulated with the FS-2 potential. Green represent $\langle 111 \rangle$ dumbbells and dislocation loops, red represent $\langle 110 \rangle$ dumbbells, and violet indicate vacancies. The figure illustrates the formation of $\tfrac{1}{2}\langle 111 \rangle$ dislocation loops, consistent with the statistical trends observed in cascade simulations.
\end{document}